\documentclass[prd,eqsecnum,showpacs,twocolumn,amsfonts,amssymb]{revtex4}

\usepackage{graphicx}

\usepackage{bm}

\newcommand{\beq}{\begin{equation}}
\newcommand{\eeq}{\end{equation}}
\newcommand{\beqs}{\begin{eqnarray}}
\newcommand{\eeqs}{\end{eqnarray}}

\begin{document}

\title{Patterns of Dynamical Gauge Symmetry Breaking}

\author{Ning Chen}
%\thanks{thomas.ryttov@stonybrook.edu}

\author{Thomas A. Ryttov}
%\thanks{thomas.ryttov@stonybrook.edu}

\author{Robert Shrock}
%\thanks{robert.shrock@stonybrook.edu}

\affiliation{
C. N. Yang Institute for Theoretical Physics \\
Stony Brook University \\
Stony Brook, NY 11794}

\begin{abstract}

We construct and analyze theories with a gauge symmetry in the ultraviolet of
the form $G \otimes G_b$, in which the vectorial, asymptotically free $G_b$
gauge interaction becomes strongly coupled at a scale where the $G$ interaction
is weakly coupled and produces bilinear fermion condensates that dynamically
break the $G$ symmetry.  Comparisons are given between Higgs and dynamical
symmetry breaking mechanisms for various models.

\end{abstract}


\maketitle

\section{Introduction} 

An outstanding question at present concerns the origin of electroweak symmetry
breaking (EWSB), in which the electroweak gauge symmetry of the Standard Model
(SM), based on the gauge group $G_{EW} = {\rm SU}(2)_L \otimes {\rm U}(1)_Y$,
where SU(2)$_L$ and U(1)$_Y$ are the factor groups for weak isospin and
hypercharge, is broken to the electromagnetic U(1)$_{em}$ subgroup. The
Standard Model hypothesizes that this symmetry breaking is due to the vacuum
expectation value (VEV) of a fundamental Higgs field that transforms as $T=1/2$
and $Y=1$. Similarly, the minimal supersymmetric Standard Model (MSSM)
attributes electroweak symmetry breaking to nonzero VEVs of the (scalar
components of) two Higgs chiral superfields with $T=1/2$ and $Y = \pm 1$. A
rather different approach is taken by technicolor (TC) theories.  In these, the
vectorial, asymptotically free technicolor gauge interaction becomes strongly
coupled at the TeV scale, producing condensates of technifermions that break
$G_{EW}$ to U(1)$_{em}$.  Other possibilities have also been studied, such as
electroweak symmetry breaking due to boundary conditions on gauge fields in
higher dimensions.  Experiments at the Large Hadron Collider (LHC) are
currently underway to answer the question of the origin of electroweak symmetry
breaking.

 In general, a comparative study of Higgs-type and dynamical approaches to the
breaking of gauge symmetries gives insights into both of these approaches.  In
this paper we shall carry out such a study.  We shall consider a class of gauge
theories with a direct-product gauge symmetry of the Lagrangian, of the form
\beq
G_{UV} = G \otimes G_b \ , 
\label{guv}
\eeq
such that as the theory evolves from some high energy scale to lower energies,
the $G_b$ interaction becomes strongly coupled at a scale $\Lambda_b$, where
the $G$ interaction is weakly coupled, and produces bilinear fermion
condensates that transform as nonsinglets under $G$ and hence dynamically break
the $G$ symmetry to a subgroup $H \subset G$, i.e., 
\beq
G \to H \quad {\rm induced \ by} \ G_b \ . 
\label{gh}
\eeq
(The subscript $b$ on $G_b$ and $\Lambda_b$ refers to their roles in the
breaking of $G$.)  The condition that the $G$ interaction is weakly coupled at
the scale $\Lambda_b$ is similar to the fact that the electroweak interaction
is weakly coupled at the scale $2^{-1/4}G_F^{-1/2} \simeq 250$ GeV where it is
broken.  However, our study is not an attempt to construct a semi-realistic
theory of dynamical EWSB, but instead focuses on gaining insights into the
differences between Higgs-type and dynamical symmetry breaking through
comparative analyses of various models.

In order for the dynamical symmetry in Eq. (\ref{gh}) to occur, the following
conditions are necessary and are therefore assumed here: (i) the $G_b$ gauge
interaction is asymptotically free, so that the running coupling $\alpha_b(\mu)
= g_b(\mu)^2/(4\pi)$ increases as the reference energy scale $\mu$ decreases;
(ii) $G_b$, considered by itself, is a vectorial gauge symmetry, so that it
does not self-break when it forms condensates, but instead remains exact; and
(iii) the content of fermions that are nonsinglets under $G_b$ is sufficiently
small so that as the $G_b$ interaction evolves from the ultraviolet to lower
energy scales, $\alpha_b(\mu)$ increases sufficiently to exceed the critical
value for the formation of the requisite $G$-breaking fermion condensates
rather than evolving in a chirally symmetric manner.  We consider several types
of symmetries $G$, both vectorial and chiral, and of both direct-product and
(semi)simple type. Although $G_b$, considered by itself (with the $G$
interaction turned off), is vectorial, the full gauge symmetry $G_{UV}$ is
chiral in all of the cases that we consider.  The $G_{UV}$ symmetry thus
requires that the fermions that are nonsinglets under $G_b$ have zero intrinsic
masses.

One can generalize the analysis further to deal with gauge symmetries
of the form
\beq
G_{UV} = G \otimes [\prod_{i=1}^k G_{b_i}] \ , 
\label{guvmultiple}
\eeq
where $k$ strongly coupled gauge interactions $G_{b_i}$, $1 \le i \le k$, play
a role in the dynamical breaking of $G$. We will focus on the simplest case,
$k=1$, but will also comment on models with $k=2$.

This paper is organized as follows.  In Sect. II we review two illustrative
examples of the type of induced gauge symmetry breaking that we consider.  In
Sect. III we carry out a comparative study of the breaking of an SU(3) gauge
symmetry to SU(2) by a Higgs field in the fundamental representation and by a
dynamical mechanism. We also discuss how color SU(3)$_c$ would be broken in a
modified Standard Model with a strongly coupled SU(2)$_L$ interaction.  In
Sect. IV we carry out a comparative study of the breaking of an SU(3) gauge
symmetry by a Higgs field in the adjoint representation and by a dynamical
mechanism.  This is generalized to SU($N$) in Sect. V.  Some further discussion
and our conclusions are given in Sects. VI and VII.

\section{Some Examples of Induced Dynamical Symmetry Breaking}

\subsection{QCD Breaking Electroweak Symmetry}

As background for our work, we first briefly review two examples of induced
dynamical symmetry breaking of weakly coupled gauge symmetries by strongly
coupled gauge interactions.  In addition to the physics that is responsible for
the main electroweak symmetry breaking at the scale $\sim 250$ GeV, there is
another source of EWSB, albeit at a much smaller mass scale.  This is quantum
chromodynamics (QCD). The color SU(3)$_c$ gauge interaction produces bilinear
quark condensates at a scale $\Lambda_{QCD} \sim 250$ MeV, in the most
attractive channel (MAC) $3 \times \bar 3 \to 1$, of the form $\langle \bar q q
\rangle = \langle \bar q_L q_R \rangle + h.c.$ Because these quark condensates
transform as weak $T=1/2$, $|Y|=1$ quantities, they break $G_{EW}$ to
electromagnetic U(1)$_{em}$.  Indeed, one could imagine a hypothetical world in
which the electroweak symmetry were not broken at the normal scale, but instead
remained valid all the way down to the QCD scale.  In this world (assuming that
the SU(3)$_c$, SU(2)$_L$, and U(1)$_Y$ running gauge couplings had
approximately their usual values), QCD would be the main source of EWSB
\cite{tc,smr}.  Such a theory would be of the form of Eqs. (\ref{guv}) and
(\ref{gh}), with
\beq
G = G_{EW} \ , \quad G_b = {\rm SU}(3)_c \ , \quad 
H = {\rm U}(1)_{em} \ . 
\label{guv_ggb_qcd}
\eeq
In this hypothetical world the $W$ and $Z$ would pick up masses given by
$m_W^2 = g^2 f_\pi^2/4$ and $m_Z^2 = (g^2 + g'^2)f_\pi^2/4$, where $g$ and $g'$
are the SU(2)$_L$ and U(1)$_Y$ running gauge couplings at the scale
$\Lambda_{QCD}$, and $f_\pi$ is the pion decay constant.

\subsection{Electroweak Symmetry Breaking by Technicolor}

Technicolor models embody the idea of dynamical electroweak symmetry breaking
\cite{tc} (recent reviews include \cite{tcrev}). In these models, the gauge
symmetry that is broken is (the electroweak part of) the SM gauge group
$G=G_{SM} = {\rm SU}(3)_c \otimes {SU}(2)_L \otimes {\rm U}(1)_Y$.  At the
scale where $G_{SM}$ is broken, all of the three gauge interactions
corresponding to its factor groups are weakly coupled. The technicolor gauge
interaction is associated with the group $G_b=G_{TC}$. Typically, $G_{TC} =
{\rm SU}(N_{TC})$ with some value of $N_{TC}$ such as 2, so these models can be
described in the notation of Eqs. (\ref{guv}) and (\ref{gh}) by
\beq
G = G_{SM} \ , \quad G_b = {\rm SU}(N_{TC}) \ , \quad 
H = {\rm SU}(3)_c \otimes {\rm U}(1)_{em} \ . 
\label{guv_tc}
\eeq
The (vectorial, asymptotically free) technicolor gauge interaction produces
condensates of technifermions $\langle \bar F F \rangle = \langle \bar F_L F_R
\rangle + h.c.$ that transform as weak $T=1/2$, $|Y|=1$ and hence break
$G_{EW}$ to U(1)$_{em}$, as indicated in Eq. (\ref{guv_tc}).  Technicolor
models are embedded in extended technicolor (ETC) in order to communicate the
electroweak symmetry breaking to the quarks and leptons.  These TC/ETC theories
are subject to a number of constraints from induced flavor-changing neutral
processes, precision electroweak data, and limits on pseudo-Nambu-Goldstone
bosons (PNGBs).

Technicolor models can be classified into two generic types: (i) one-family
models, in which the technifermions comprise one SM family, and (ii)
one-doublet models, in which, among the technifermions, there is only a single
electroweak doublet. One-family (but not one-doublet) technicolor models
feature a color-octet technivector meson resonance, as well as a
color-nonsinglet PNGB's.  Many searches for technihadrons have been carried out
\cite{tcrev}.  Recent LHC results from the ATLAS and CMS experiments have
set lower limits of order 1.5 TeV on a color-octet technivector meson
\cite{atlas,cms}.

\section{Breaking an SU(3) Gauge Symmetry to SU(2)}

In this section we shall compare Higgs and dynamical mechanisms for breaking an
SU(3) gauge symmetry to SU(2).  We assume that the fermion content of the
theory is such that the fermions that are only nonsinglets under SU(3) form a
vectorlike sector.  We shall begin by considering an abstract asymptotically
free SU(3) theory at a sufficiently high scale that it is weakly coupled.

\subsection{Higgs Mechanism to break SU(3) to SU(2) }

The requisite breaking can be accomplished by including a Higgs field $\phi$
that transforms as a fundamental (triplet) representation of the SU(3) group,
with a potential
\beq
V = \frac{\mu^2}{2} \phi^\dagger \phi + 
\frac{\lambda}{4} ( \phi^\dagger \phi)^2 \ , 
\label{vsu3higgs}
\eeq
where $\mu^2 < 0$ (and $\lambda > 0$ in order for $V$ to be bounded from
below).  This potential is minimized for a nonzero value of the $\phi$ vacuum
expectation value.  Without loss of generality, one can use the SU(3) gauge
invariance to define directions in SU(3) space so that this has the form
$\langle \phi \rangle = (0,0,1)^T \, v$, and one can perform a global phase
redefinition on $\phi$ to make $v$ real.  This breaks SU(3) to the SU(2)
subgroup generated by the the SU(3) generators $T_a$ with $a=1,2,3$ (in the
usual Gell-Mann ordering of these generators).  Of the six real components of
the $\phi$ field, five are Nambu-Goldstone bosons and are absorbed by the five
gauge bosons in the coset space ${\rm SU}(3)/{\rm SU}(2)$ to form the
longitudinal polarization states of the resultant massive vector bosons.  The
resultant vector boson masses are $\propto g_3(v)v$, where $g_3 \equiv
g_3(\mu)$ is the running SU(3) gauge coupling at the scale $\mu = v$.  The
sixth component of the $\phi$ field forms a physical Higgs boson with a mass
$\sim \sqrt{\lambda} \, v$.  This is a singlet under the residual SU(2) gauge
interaction.

As noted above, we assume that this breaking occurs at a scale $v$ that is
large compared with the scale where the running SU(3) gauge coupling
$\alpha_3(\mu) = g_3(\mu)^2/(4\pi)$ would have grown to O(1) and the theory
would thus have become strongly coupled.  This assumption is necessary for this
model to fall under the class of theories that are considered in this paper. If
one were to relax this assumption, the analysis would become more complicated,
because one would not be able to perform a perturbative analysis of the Higgs
sector.  (For a recent discussion of this strongly coupled case and further
references, see \cite{mg}).  Below the scale $v$, the resultant SU(2) theory
would have a fermion sector consisting of the SU(2)-nonsinglet components of
the original SU(3) fermion sector, together with the SU(2) gluons, with a gauge
coupling inherited from the original SU(3) theory.  This SU(2) theory would
then evolve further into the infrared.  With a sufficiently small fermion 
content $\{ f \}$, the SU(2) coupling would eventually increase 
to O(1) at a lower scale $\Lambda_2$, where the SU(2) interaction would 
confine and produce bilinear fermion condensates.  There would thus be a
spectrum of SU(2)-singlet meson and (bosonic) baryons, together with 
glueballs (which would mix with the mesons to produce mass eigenstates) at this
lower scale $\Lambda_2$.

There are several properties of this Higgs mechanism that will be contrasted
with the induced dynamical breaking mechanism to be discussed next.  First,
{\it a priori}, one has the freedom to choose the coefficient $\mu^2$ in the
Higgs potential (\ref{vsu3higgs}) to be positive or negative.  Since one wants
to construct the Higgs mechanism to break SU(3), one chooses $\mu^2 < 0$, but
this sign choice could be considered to be {\it ad hoc}, since one does not
give any deeper explanation for this choice.  Second, the Higgs mechanism
predicts physical pointlike Higgs particle(s), whereas in a dynamical
mechanism, although the $G_b$ interaction leads to various $G_b$-singlet bound
states, including some with angular momentum $J=0$, the properties of these
states are not, in general, the same as those of a Higgs particle. Third, as
is well known, this potential is unstable to large loop corrections and is thus
sensitive to the nature of the ultraviolet completion of the theory (i.e., has
a hierarchy problem).  A fourth and related point is that the Higgs sector is
not asymptotically free, i.e., the beta function for the quartic coupling,
$d\lambda/dt$, is positive, where $t = \ln \mu$. Because of this, if one fixes
$\lambda$ at the scale $v$, say, then one must worry about a possible Landau
pole in $\lambda$ that could occur at a scale $\mu >> v$.  An equivalent way to
phrase this is that if one fixes $\lambda$ at a high scale in the ultraviolet,
then $\lambda$ decreases as $\mu$ decreases and is subject to an upper bound at
a much lower scale such as $v$ \cite{triviality}.

\subsection{Induced Dynamical Breaking of SU(3) to SU(2) }

In this subsection we discuss how one can produce the breaking of the SU(3)
symmetry to SU(2) in a dynamical manner.  For $G_b$ we choose the smallest
non-Abelian Lie group, SU(2)$_b$, so that $G_{UV} = {\rm SU}(3) \otimes {\rm
SU}(2)_b$, in the notation of Eq. (\ref{guv}). To the set of fermions $\{f\}$ 
transforming vectorially under SU(3) we add the following chiral fermions
(where $a$ and $\alpha$ denote SU(3) and SU(2)$_b$ gauge indices, respectively
and the numbers in parenthese denote the dimensionalities of the
representations of $G_{UV}$ ): (i) $\zeta^{a \alpha}_L \ : \ (3,2)$; (ii)
$\eta^\alpha_L \ : \ (1,2)$; and (iii) $\chi^a_{p,R} \ : \ (3,1)$ with
$p=1,2$.  This set of fermions is similar to the set that one of us used in
Ref. \cite{mg}. Since the SU(2)$_b$ gauge interaction is asymptotically free,
as the reference energy scale $\mu$ decreases from large values, the running
coupling $\alpha_b(\mu)$ increases.  The SU(2)$_b$-nonsinglet fermions comprise
four chiral Weyl fermions or equivalently, two Dirac fermions.  This is well
below the estimated critical number $N_{f,cr} \sim 8$ beyond which the
SU(2)$_b$ theory would evolve into the infrared in a chirally symmetric manner
\cite{chipt}.  Therefore, we can conclude that as $\mu$ decreases to the
scale $\mu=\Lambda_b$ such that $\alpha_b(\mu) \sim O(1)$, the SU(2)$_b$
interaction produces bilinear fermion condensates.  The most attractive channel
for the fermion condensation is $2 \times 2 \to 1$.  One such condensate is of
the form $\langle \epsilon_{\alpha\beta} \, \zeta^{a \alpha \ T}_L C \,
\zeta^{b \beta}_L \rangle$, where $\epsilon_{\alpha\beta}$ is the antisymmetric
tensor density for SU(2)$_b$. This is automatically antisymmetrized in the
SU(3) indices $a,b$ and hence is proportional to
\beq
\langle \epsilon_{abc} \epsilon_{\alpha\beta} \, \zeta^{a \alpha \ T}_L C \,
\zeta^{b \beta}_L \rangle \ , 
\label{zetazeta}
\eeq
where $\epsilon_{abc}$ is the antisymmetric tensor density for SU(3).  The
condensate (\ref{zetazeta}) transforms as conjugate fundamental ($\bar 3$)
representation of SU(3) and therefore dynamically breaks SU(3) to an SU(2)
subgroup.  A second condensate formed by the SU(2)$_b$ interaction is
\beq
\langle \epsilon_{\alpha\beta} \, \zeta^{a \alpha \ T}_L C \eta^{\beta}_L 
\rangle \ .
\label{zetaeta}
\eeq
This transforms as a fundamental representation of SU(3) and hence also
breaks it to an SU(2) subgroup.  One can use vacuum alignment arguments
\cite{vacalign} to infer that these SU(2) subgroups are the same.  Then,
without loss of generality, one may choose the index $c=3$ in the condensate
(\ref{zetazeta}) and $a=3$ in the condensate (\ref{zetaeta}).  The residual
SU(2) subgroup preserved by these condensates is thus the one generated by
$T_a$, $a=1,2,3$ in SU(3). The fermions $\zeta^{a \ \alpha}_L$ and
$\eta^\alpha_L$ with $a=1,2,3$, $\alpha=1,2$ involved in these condensates gain
dynamical masses of order $\Lambda_b$ and are integrated out of the low-energy
effective field theory that is operative as scales $\mu < \Lambda_2$.  The two
copies of $\chi^a_{p,L}$ decompose as two doublets under the resultant SU(2)
for $a=1,2$ (while the $a=3$ components form two singlets).  The fermion
content of this low-energy SU(2) theory thus consists of these two doublets,
together with the SU(2)-nonsinglet components of the set $\{f\}$.  With an
asymptotically free SU(2), the coupling $\alpha_2(\mu)$, which is inherited
from the weakly coupled SU(3) theory will increase as $\mu$ decreases below
$\Lambda_b$, and if the fermion content is sufficiently small so that
$\alpha_2(\mu)$ grows to O(1) at a lower scale $\Lambda_2$, the SU(2) gauge
interaction will confine and form bilinear fermion condensates at this scale.
Given that $\alpha_2$ is small at the scale $\Lambda_b$ and increases only
logarithmically, it follows that $\Lambda_2 << \Lambda_b$.

If one were to turn off the SU(3) gauge interaction, the SU(2)$_b$ theory would
have a classical U(4) or equivalently ${\rm SU}(4) \otimes {\rm U}(1)$ global
chiral symmetry.  The U(1) is broken by SU(2)$_b$ instantons, so that the
actual non-anomalous global chiral symmetry would be the SU(4) (generated by
global transformations of the $\zeta^{a \ \alpha}_L$ and $\eta^\alpha_L$ among
each other for a fixed $\alpha$.)  The bilinear condensates would break this to
Sp(4), with the resultant appearance of five Nambu-Goldstone bosons.  Turning
on the SU(3) gauge interaction explicitly breaks this global symmetry, although
the breaking is weak, since $\alpha_3$ is small.

We contrast this dynamical breaking with the corresponding Higgs mechanism
presented above. First, the dynamical mechanism is more predictive, in
the sense that once one has specified the gauge interaction $G_b$ and the
fermion content, the resulting fermion condensation and symmetry breaking
follow automatically; one does not have to make an {\it ad hoc} choice of a
parameter, as one does with the choice $\mu^2 < 0$ in the Higgs
potential (\ref{vsu3higgs}).  Second, the theory does not suffer from a
hierarchy problem, i.e., is not sensitively dependent on an ultraviolet
completion, in contrast to the Higgs mechanism.  Third, by construction, both
the SU(3) and the SU(2)$_b$ sectors are asymptotically free, again in contrast
with the Higgs mechanism, in which the quartic coupling is not asymptotically
free.

\subsection{Induced Breaking of SU(3)$_c$ in a Modified Standard Model}

Here we discuss another way to break an SU(3) gauge symmetry dynamically.  In
this case we will take the SU(3) to be the color SU(3)$_c$ group of the
Standard Model.  The point here is that with a modification of the properties
of the Standard Model, color SU(3)$_c$ would, in fact, be dynamically broken by
the SU(2)$_L$ gauge interaction.  Our analysis also addresses a fundamental
question in particle physics.  One of the profound properties of nature is the
fact that it is the chiral part of $G_{SM}$ that is broken, leaving as a
residual exact subgroup a symmetry that is vectorial, namely $H = {\rm SU}(3)_c
\otimes {\rm U}(1)_{em}$.  This is naturally explained in the Standard
Model Higgs mechanism and also in technicolor theories.  One is led, then, to
ask how general this property is in quantum field theory.  That is, can one
construct a model that exhibits dynamical breaking of a vectorial gauge
symmetry?  Clearly, this requires more than one gauge interaction to be
present, since if one has just a single vectorial gauge interaction and it
becomes strongly coupled and produces condensates, then the most attractive
channel is $R_i \times \bar R_i \to 1$ for the one or more fermion
representations $R_i$ in the theory, so it does not self-break
\cite{deltac2,selfbreak}.

Let us thus consider a theory with the same gauge group, $G_{SM}$, but make
two changes: (i) first, we remove the usual breaking of SU(2)$_L$ at the 250
GeV scale, and (ii) we arrange the values of the gauge couplings so that at a
scale $\Lambda_2$ considerably larger than $\Lambda_{QCD}$, where SU(3)$_c$
(and U(1)$_Y$) are weakly coupled, the SU(2)$_L$ interaction becomes
strongly coupled, with $\alpha_2(\Lambda_2) = g(\Lambda_2)^2/(4\pi)$ of order
unity.  The SU(2)$_L$ sector contains $N_{gen}(N_c+1)=12$ chiral fermion
doublets (where $N_{gen}$ denotes the number of SM fermion generations), so
that the SU(2)$_L$ gauge interaction is asymptotically free, with leading
coefficient
\beq
(b_1)_{SU(2)_L} = \frac{1}{3}[22-(N_c+1)N_{gen}] \ . 
\label{b1su2l}
\eeq
Given that there is no breaking of $G_{EW}$, the fermions are massless, so they
all contribute to the SU(2)$_L$ beta function. To illustrate this dynamical
breaking in the simplest context, we assume $N_{gen}=1$, so that there are four
chiral SU(2)$_L$ doublets, or equivalently, $N_f=2$ Dirac doublets.  This is
well within the phase in which SU(2)$_L$ confines and spontaneously breaks
global chiral symmetry. The model thus contains one family of SM fermions:
$Q^{ai}_L = {u^a \choose d^a}_L$, $u^a_R$, $d^a_R$, $L_L^i={\nu_e \choose
e}_L$, and $e_R$, where $a$ and $i$ are SU(3)$_c$ and SU(2)$_L$ gauge indices,
respectively.

The most attractive channel for the strongly coupled SU(2)$_L$ interaction is
$2 \times 2 \to 1$, and it produces several condensates in this channel.  The
first of these is of the form $\langle \epsilon_{k \ell} Q^{a,k \ T}_L C
Q^{b,\ell}_L \rangle$, where $\epsilon_{k \ell}$ is the antisymmetric tensor
density for SU(2)$_L$.  This is automatically antisymmetric in SU(3)$_c$
indices and hence is proportional to
\beq
\langle \epsilon_{abc} \epsilon_{k \ell } Q^{a,k \ T}_L C
Q^{b,\ell}_L \rangle =
2\langle \epsilon_{abc} u^{a \ T}_L C d^b_L \rangle \ ,
\label{qq_condensate}
\eeq
This transforms as a $(3 \times 3)_{as} = \bar 3$ under SU(3)$_c$ (where the
subscript $as$ stands for antisymmetric) and hence breaks SU(3)$_c$ to a
subgroup SU(2)$_c$. It also breaks U(1)$_Y$.  As is clear from the fact that
electric charge satisfies $Q=T_3+(Y/2)$ and the fact that the condensate
(\ref{qq_condensate}) is invariant under SU(2)$_L$, it also violates electric
charge invariance. Without loss of generality, we choose the breaking direction
of SU(3)$_c$ as the third direction, so that the $u^a_L$ and $d^a_L$ quarks
with color indices $a=1, \ 2$ occur in the condensate (\ref{qq_condensate}) and
hence gain dynamical masses of order $\Lambda_2$.

The strong SU(2)$_L$ interaction would also produce the condensate
\beq
\langle \epsilon_{k \ell } Q^{a,k  \ T}_L C L^{\ell}_L \rangle \ = 
\langle u^{a \ T}_L C e_L - d^{a \ T}_L C \nu_{e,L} \rangle \ . 
\label{ql_condensate}
\eeq
This also breaks SU(3)$_c$ to an SU(2)$_c$ subgroup and violates hypercharge
and electric charge.  As in our discussion above, a vacuum alignment argument
can be used to infer that the condensate (\ref{ql_condensate}) breaks SU(3)$_c$
to the same SU(2)$_c$ as the condensate (\ref{qq_condensate}), so that the
color index in Eq. (\ref{ql_condensate}) has the value $a=3$. This SU(2)$_c$ is
the one generated by the color generators $(T_a)$ with $a=1,2,3$.  Thus, this
model is of the form in Eqs. (\ref{guv}) and (\ref{gh}) with
\beqs
& & G = {\rm SU}(3)_c \otimes {\rm U}(1)_Y  \ , \quad 
G_b = {\rm SU}(2)_L \ , \quad H = {\rm SU}(2)_c \ . 
\cr\cr 
& & 
\label{guv_su2l}
\eeqs
In addition to breaking these gauge symmetries, the condensate
(\ref{qq_condensate}) breaks baryon number by $\Delta B = 2/3$, while the
condensate (\ref{ql_condensate}) breaks $B$ by $\Delta B = 1/3$ and lepton
number $L$ by $\Delta L=1$.  The quarks $u^a_L$, $d^a_L$ with $a=1,2,3$ and the
leptons $e_L$, and $\nu_{e,L}$ involved in these condensates gain dynamical
masses of order $\Lambda_2$. (The actual mass eigenstates involve linear
combinations of these fields.) Similarly, the five gluons in the coset ${\rm
SU}(3)_c/{\rm SU}(2)_c$ corresponding to the broken generators of SU(3)$_c$
gain dynamical masses of order
\beq
m_g \sim g_3(\Lambda_2)\Lambda_2
\label{mg}
\eeq
and the abelian U(1)$_Y$ gauge boson $B$ gains a mass
\beq
m_B \sim g'(\Lambda_2)\Lambda_2 \ . 
\label{mb}
\eeq
Since by our assumptions, SU(3)$_c$ and U(1)$_Y$ are weakly coupled
at this scale, the masses of these five gluons and of the one $B$ boson are
smaller than the dynamically produced fermion masses.  

Of the quarks and leptons in this $N_{gen}=1$ model, all of the components of
the $N_c+1=4$ SU(2)$_L$ doublets are involved in the condensates
(\ref{qq_condensate}) and (\ref{ql_condensate}) and gain dynamical masses of
order $\Lambda_2$.  These fermions are thus integrated out of the low-energy
effective theory below $\Lambda_2$.  The SU(2)$_c$ gauge symmetry of this
low-energy effective field theory remains exact.  The content of nonsinglet
fermions in this low-energy theory consists of $u^a_R$ and $d^a_R$ with
$a=1,2$, which form two Weyl fermions, or equivalently, one Dirac fermion.  The
SU(2)$_c$ gauge coupling $\alpha_{2c}$ is inherited from the SU(3)$_c$ theory
and is small at $\Lambda_2$, but eventually grows to O(1) at a much lower
$\Lambda_{2c} << \Lambda_2$.  At this lower scale $\Lambda_{2c}$, the SU(2)$_c$
theory confines and produces a bilinear fermion condensate,
\beq
\langle \epsilon_{ab} \, u^{a \ T}_R C d^b_R \rangle \ , 
\label{udrcondensate}
\eeq
where here $\epsilon_{ab}$ is the antisymmetric tensor density of SU(2)$_c$.
This SU(2)$_c$ theory has a classical U(2), or equivalently, ${\rm SU}(2)
\otimes {\rm U}(1)$ global chiral symmetry defined by transformations that mix
up the $u_R^a$ and $d_R^a$ fields (for fixed $a$). The U(1) is broken by
SU(2)$_c$ instantons, so that the actual non-anomalous global chiral symmetry
is SU(2). In general, an SU(2) gauge theory with $N_d$ massless chiral Weyl
fermions transforming according to the fundamental representation (with
$N_d=2k$ even to avoid a global Witten anomaly) has an SU($2k$) global chiral
symmetry corresponding to transformations that mix up the $2k$ chiral doublets.
Formation of condensates involving these doublets breaks this global symmetry
to Sp($2k$).  Since the orders of these groups are $4k^2-1$ and $k(2k+1)$,
respectively, this entails the breaking of $2k^2-k-1$ generators of SU($2k$),
and the resultant appearance of this number of massless Nambu-Goldstone
bosons. In this SU(2)$_c$ theory, there are $N_d=2$ chiral fermions, i.e.,
$k=1$, so the SU(2) global chiral symmetry is equivalent to Sp(2), and there is
no chiral symmetry breaking due to the formation of the condensate
(\ref{udrcondensate}).

It is also worthwhile to comment on the situation concerning global chiral
symmetry at the higher scale, above $\Lambda_2$. In the present model with its
one generation of SM fermions, if one turns off the SU(3)$_c$ and U(1)$_Y$
couplings, then, an an energy above $\Lambda_2$, the SU(2)$_L$ theory has a
non-anomalous global SU($N_d$) symmetry, where $N_d=N_c+1=4$.  The condensates
(\ref{qq_condensate}) and (\ref{ql_condensate}) break this to Sp(4), leading to
the appearance of five Nambu-Goldstone bosons. Since the NGBs couple in a
derivative manner, their scattering amplitudes are suppressed by powers of
center-of-mass energy $\sqrt{s}/\Lambda_2$ and hence they are progressively
more weakly coupled as the energy scale decreases further below $\Lambda_2$
\cite{pngb}.  Turning on the SU(3)$_c$ and U(1)$_Y$ couplings explicitly breaks
the global SU(4) symmetry, but also the would-be NGBs are absorbed to form the
longitudinal components of the five vector bosons in the coset ${\rm
SU}(3)_c/{\rm SU}(2)_c$. This process is reminiscent of the mechanism by which
technicolor gives masses to the $W$ and $Z$ bosons.

Our analysis here answers the question that we posed at the beginning of this
subsection concerning the breaking of a vectorial, as contrast to a chiral,
gauge symmetry.  Our answer is that it is certainly possible for a vectorial
gauge symmetry to be broken, if this breaking is induced by another strongly
coupled interaction.  The reason that SU(2)$_L$ does not break SU(3)$_c$ in
nature is a consequence of the fact that SU(2)$_L$ is broken well above the
scale where its coupling would have become large enough to produce the
condensates (\ref{qq_condensate}) and (\ref{ql_condensate}).  The resultant $W$
and $Z$ are massive and weakly coupled and their interactions are too weak to
induce such condensates.  Indeed, even if the SU(2)$_L$ symmetry were not
broken at this higher scale, it would be broken by the quark condensates at the
QCD scale, as discussed above, before it could become strong enough to break
SU(3)$_c$.

\section{Induced Dynamical Breaking of a Gauge Symmetry by Adjoint Fields: 
an Illustrative Model with $G={\rm SU}(3)$}

\subsection{Higgs Mechanism} 

We next consider induced breaking of a gauge symmetry by fields that transform
as the adjoint representation of the gauge group.  In this section we discuss
SU(3) because of some special properties that it has, and in the next section
we discuss SU($N$) for general $N \ge 4$.  We begin by constructing a Higgs
mechanism for this breaking. We assume that the theory contains a Higgs field
$\phi$ transforming according to the adjoint (i.e., octet) representation of
SU(3), with an appropriate Higgs potential. We will write the components of
$\phi$ as $\phi^i_j$, $1 \le i,j \le 3$; these are subject to the trace
condition ${\rm Tr}(\phi) = \sum_{i=1}^3 \phi^i_i = 0$.  In general, when using
the adjoint representation of SU($N$), in addition to the notation $\phi^i_j$
with $1 \le i,j \le N$, it will also be convenient to use an equivalent
notation $\phi_a$, with $1 \le a \le N^2-1$, that indicates the 1--1
correspondence with the $N^2-1$ generators $T_a$ of SU($N$).  Thus the
$\phi^i_j$ form the entries of a matrix given by $\sqrt{2} \,
\sum_{a=1}^{N^2-1} \phi_a T_a$.  

We will require that the Higgs part of the Lagrangian be invariant under the
replacement $\phi \to -\phi$.  It follows that the Higgs potential contains
only quadratic and quartic terms in $\phi$.  For a general SU($N$) theory with
a Higgs field in the adjoint representation, there are two independent quartic
terms, proportional to $[{\rm Tr}(\phi^2)]^2$ and ${\rm Tr}(\phi^4)$.  For the
special values $N=2$ or $N=3$, $[{\rm Tr}(\phi^2)]^2=2{\rm Tr}(\phi^4)$, so
there is only one independent quartic term. For the present case of SU(3), the
Higgs potential may thus be written as
\beq 
V = \frac{\mu^2}{2}{\rm Tr}(\phi^2) + \frac{\lambda}{4}[{\rm Tr}(\phi^2)]^2 \
. 
\label{vsu3}
\eeq
Here we take $\mu^2 < 0$ to get the symmetry breaking.  This potential is
minimized for a Higgs field vacuum expectation value (VEV) of the form
\beq
\langle \phi \rangle = T_8 \, v 
\label{su3higgsvev}
\eeq
where $v$ can be made real by a global rephasing of $\phi$ and $T_8$ is the
second member of the Cartan subalgebra of SU(3),
\beq
T_8 = \frac{1}{2\sqrt{3}} 
       \left( \begin{array}{ccc}
         1 & 0 & 0 \\
         0 & 1 & 0 \\
         0 & 0 & -2   \end{array} \right )
\label{t8su3}
\eeq
The VEV (\ref{su3higgsvev}) breaks SU(3) according to the pattern
\beq
{\rm SU}(3) \to {\rm SU}(2) \otimes {\rm U}(1) \ . 
\label{su3tosu2xu1}
\eeq
Since ${\rm SU}(3)$ has order eight, while ${\rm SU}(2) \otimes {\rm U}(1)$ has
order four, there are four broken generators of SU(3), namely the $T_a$ with
$a=4,5,6,7$ in the standard Gell-Mann basis.  The corresponding components
$\phi_a$ are Nambu-Goldstone bosons and are absorbed to become the longitudinal
components of the massive vector bosons. Four physical Higgs fields remain,
with masses $\sim \sqrt{\lambda} \, v$.  Of these, $\phi_a$, $a=1,2,3$
transform as the adjoint representation of the residual SU(2) gauge interaction
and, assuming that it confines, they are thus confined in SU(2)-singlet bound
states.  Since we have assumed that the SU(2) theory is weakly coupled at the
scale $\mu=v$, and since its coupling increases only logarithmically, the SU(2)
confinement scale $\Lambda_2$ is much smaller than $v$.  In passing, we note
that although a Higgs VEV of the form $\phi = {\rm diag}(a,b,-(a+b))$, with
$|a| \ne |b|$ is, {\it a priori}, possible, and would break SU(3) to U(1)
rather than ${\rm SU}(2) \otimes {\rm U}(1)$, it does not minimize the Higgs
potential.

\subsection{Dynamical Breaking Mechanism with Adjoint Fields}

To study the dynamical breaking of the SU(3) symmetry by an SU($N_b$) gauge
interaction, we must choose a value of $N_b$ and a requisite sector comprised
of one or more fermion fields that transform as nonsinglets under both $G={\rm
SU}(3)$ and $G_b={\rm SU}(N_b)$.  For our model we choose a chiral fermion that
transforms as an adjoint of SU(3) and a fundamental representation of
SU($N_b$):
\beq
(\psi^i_{j,L})^\alpha \ : \quad (8,N_b) \ , 
\label{psisu3}
\eeq
where the numbers in parentheses are the dimensions of the representation under
$G_{UV} = {\rm SU}(3) \otimes {\rm SU}(N_b)$, the indices $i,j$ are SU(3)
indices, and $\alpha=1,...,N_b$ is an SU($N_b$) index. Because
$(\psi^i_{j,L})^\alpha$ transforms according to a self-adjoint representation
of SU(3), it does not contribute any gauge anomaly to the SU(3) theory.  We
take $N_b=2$, the minimal value, so $G_b = {\rm SU}(2)_b$.  As above,
we will also use the equivalent notation $\phi_{a,L}$, $1 \le a \le 8$.  The
$(\psi^i_{j,L})^\alpha$ form the components of a matrix given by $\sqrt{2} \,
\sum_{a=1}^8 \psi_{a,L}^\alpha T_a$.

The SU(2)$_b$ gauge interaction is asymptotically free, with the leading beta
function coefficient $b_1=14/3$ (see Appendix II for notation).  The fermion
$(\psi^i_{j,L})^\alpha$ amounts to 8 Weyl doublets, or equivalently, 4 Dirac
doublets, of SU(2).  Since this number is well below the estimated critical
value $N_{f,cr} \simeq 8$ separating the (zero-temperature) phase with
confinement and spontaneous chiral symmetry breaking from the chirally
symmetric phase \cite{chipt}, we can conclude that the SU(2) interaction
confines and produces bilinear condensates.  These occur in the most attractive
SU(2)$_b$ channel (MAC), which is $2 \times 2 \to 1$, with a condensate of the
form $\langle \epsilon_{\alpha\beta}(\psi^i_{j,L})^{\alpha \ T} C
(\psi^k_{\ell,L})^\beta \rangle$, where here $\epsilon_{\alpha\beta}$ is the
antisymmetric tensor density for SU(2)$_b$.  The $\epsilon_{\alpha\beta}$
contraction antisymmetrizes the bilinear fermion product, so that in the full
Clebsch-Gordan decomposition,
\beq
8 \times 8 = 1_s + 8_s + 8_a + 10_a + \overline{10}_a + 27_s
\label{88cg}
\eeq
(where the subscripts $s$ and $a$ denote symmetric and antisymmetric
combinations), the above condensate must be one of the antisymmetric products,
namely $8_a$ or $10_a$. We next use a vacuum alignment argument
\cite{vacalign}, according to which the symmetry breaking should preserve as
large a subgroup symmetry as possible.  Now, relative to the 
maximal subgroup ${\rm SU}(2) \otimes {\rm U}(1)$, an octet of SU(3) has the
decomposition
\beq
8_{SU(3)} = 3_0 + 2_1 + 2_{-1} + 1_0  \ , 
\label{8insu3}
\eeq
where the numbers on the right-hand side are the dimensionalities of the SU(2)
representations and the subscripts are the hypercharges with the Gell-Mann
normalization. In contrast, the decuplet has the decomposition
\beq
10_{SU(3)} = 4_1 + 3_0 + 2_{-1} + 1_{-2} \ .  
\label{10insu3}
\eeq
Of these, only the octet contains a piece that is a singlet under 
${\rm SU}(2) \otimes {\rm U}(1)$.  Using a vacuum alignment argument, we
therefore can infer that the condensate transforms as the $1_0$ piece of the
octet of SU(3) and hence has the form 
\beq
\langle \epsilon_{\alpha\beta}(\psi^i_{j,L})^{\alpha \ T} C
(\psi^j_{\ell,L})^\beta \rangle \propto (T_8)^i_\ell \, \Lambda_b^3 \ . 
\label{psipsi_su3}
\eeq
In the equivalent notation using $\psi_{a,L}^\alpha$ with $1 \le a \le 8$, the
condensate has the form $\langle \epsilon_{\alpha\beta}f_{ab8}
(\psi_{a,L})^{\alpha \ T} C (\psi_{b,L})^\beta \rangle$, where the $f_{abc}$
are the structure constants of the SU(3) Lie algebra. The nonzero structure
constants $f_{ab8}$ with $a < b$ that enter here are $f_{458}$ and
$f_{678}$. The condensate (\ref{psipsi_su3}) dynamically breaks SU(3) to ${\rm
SU}(2) \otimes {\rm U}(1)$, as in Eq. (\ref{su3tosu2xu1}).  As in the Higgs
case, there are four broken generators, namely the $T_a$ with $a=4,5,6,7$.  The
Nambu-Goldstone modes involving $\psi_{a,L}^\alpha$, $a=4,5,6,7$, contracted on
the SU(2)$_b$ indices $\alpha=1,2$ to form SU(2)$_b$ singlets, are absorbed by
the corresponding SU(3) gauge bosons, forming longitudinal polarization states
and giving them masses. The remaining $\psi_{a,L}^\alpha$ fermions with
$a=1,2,3,8$ are bound in SU(2)$_b$-singlet states. Furthermore, of these bound
states, the ones with $a=1,2,3$ transform as the adjoint representation of the
residual SU(2) gauge interaction, and, assuming that it confines, they are
confined in SU(2)-singlet bound states.  This is reminiscent of the situation
with the corresponding components of Higgs fields in the situation where one
uses a Higgs mechanism for the breaking. 

A comment is in order here concerning chiral symmetry in this model. The
$\psi_{a,L}^\alpha$ fermions have zero Lagrangian masses, and hence, if one
turns off the SU(3) gauge interaction completely, the theory has a
(non-anomalous) SU(8) global chiral symmetry.  In general, a full set of
bilinear fermion condensates breaks this to Sp(8).  As noted above, the
breaking of SU($2k$) to Sp($2k$) entails $2k^2-k-1$ broken generators and
corresponding massless Nambu-Goldstone bosons.  With $k=4$, this means 27 NGBs
in the present case.  When one turns on the SU(3) gauge coupling, this
explicitly breaks the global SU(8) chiral symmetry, and, moreover, the vacuum
alignment argument suggests which condensates form, as we have
discussed above.

\section{Induced Breaking of an SU($N$) Symmetry by Adjoint Fields}

\subsection{General}

In this section we carry out a comparative study of a Higgs mechanism versus
dynamical breaking of an SU($N$) gauge symmetry with $N \ge 4$ by fields
transforming according to the adjoint representation of this group.  Two
general types of breaking patterns of the SU($N$) symmetry will be
relevant. Both of these involve breaking to a maximal subgroup of SU($N$), with
the same rank (dimension of the Cartan subalgebra of mutually commuting
generators) as SU($N$), namely $N-1$.  However, these subgroups have different
orders (numbers of generators).  The first of these symmetry-breaking patterns
is
\beq
{\rm SU}(N) \to {\rm SU}(N-1) \otimes {\rm U}(1) \ . 
\label{symbreak_lam2neg}
\eeq
The residual symmetry group has order 
\beq
o \Big [ {\rm SU}(N-1) \otimes {\rm U}(1) \Big ] = (N-1)^2
\label{order_symbreak_lam2neg}
\eeq
so the symmetry reduction in Eq. (\ref{symbreak_lam2neg}) involves the breaking
of 
\beq
\Delta o = 2(N-1)
\label{delta_o_symbreak_lam2neg}
\eeq
generators of SU($N$), which is the dimension of the coset 
\beq
{\rm SU}(N)/[{\rm SU}(N-1) \otimes {\rm U}(1)] \ . 
\label{coset_lam2neg}
\eeq

The second type of symmetry breaking pattern leads to a residual symmetry
involving three factor groups.  To describe this, it is convenient to deal
separately with the cases of even and odd $N$. For even $N=2k$, a 
possible symmetry-breaking pattern is 
\beq
{\rm SU}(N) \to {\rm SU}(N/2) \otimes {\rm SU}(N/2) \otimes {\rm U}(1) \ .
\label{symbreak_neven_lam2pos}
\eeq
The residual symmetry group has order 
\beq
o \Big [ {\rm SU}(N/2) \otimes {\rm SU}(N/2) \otimes {\rm U}(1) \Big ] = 
\frac{N^2}{2} - 1 = 2k^2-1 \ , 
\label{order_symbreak_neven_lam2pos}
\eeq
so that (\ref{symbreak_neven_lam2pos}) involves the breaking of 
\beq
\Delta o = \frac{N^2}{2} = 2k^2
\label{deltao_symbreak_neven_lam2pos}
\eeq
generators of SU($N$). 

For odd $N=2k+1$, a possible symmetry-breaking pattern is
\beq
{\rm SU}(N)\to {\rm SU}((N+1)/2) \otimes {\rm SU}((N-1)/2) \otimes 
{\rm U}(1) \ .
\label{symbreak_nodd_lam2pos}
\eeq
The residual symmetry group has order
\beqs
& & o \Big [ {\rm SU}((N+1)/2) \otimes {\rm SU}((N-1)/2) \otimes 
{\rm U}(1) \Big ] \cr\cr
& = & \frac{N^2-1}{2} = 2k(k+1) \ , 
\label{order_symbreak_nodd_lam2pos}
\eeqs
so that (\ref{symbreak_nodd_lam2pos}) involves the breaking of 
\beq
\Delta o = \frac{N^2-1}{2} = 2k(k+1)  
\label{deltao_symbreak_nodd_lam2pos}
\eeq
generators of SU($N$).  The symmetry breaking patterns
(\ref{symbreak_neven_lam2pos}) and (\ref{symbreak_nodd_lam2pos}) can be
expressed in a unified manner as
\beq
{\rm SU}(N) \to {\rm SU}(N-\ell) \otimes {\rm SU}(\ell) \otimes {\rm U}(1) \ , 
\label{symbreak_lam2pos}
\eeq
where $\ell = [N/2]_{ip}$ and $[\nu]_{ip}$ denotes the \underline{i}ntegral
\underline{p}art of the real number $\nu$.

As we will discuss further below in the context of dynamical symmetry breaking,
a vacuum alignment argument prefers a symmetry-breaking pattern that yields the
largest residual symmetry.  The size of the subgroup that constitutes the
residual symmetry can be characterized by its rank and order.  All of the
patterns above satisfy the condition that the rank of the residual symmetry
group should be maximal, i.e., the same as that of SU($N$), namely $N-1$.
Concerning the differences in the orders of the various possible subgroups
resulting from the symmetry breaking of SU($N$), we calculate, for even $N=2k$,
the difference
\begin{widetext}
\beq
o \Big [ {\rm SU}(N-1) \otimes {\rm U}(1) \Big ] -
o \Big [ {\rm SU}(N/2) \otimes {\rm SU}(N/2) \otimes {\rm U}(1) \Big ] = 
\frac{(N-2)^2}{2} = 2(k-1)^2 \ . 
\label{dif1a}
\eeq
This difference is positive semidefinite, and positive-definite for $k \ge 2$,
i.e., $N \ge 4$. For odd $N=2k+1$,
\beq
o \Big [ {\rm SU}(N-1) \otimes {\rm U}(1) \Big ] - 
o \Big [ {\rm SU}((N+1)/2) \otimes {\rm SU}((N-1)/2) \otimes {\rm U}(1) \Big ]
 = \frac{(N-1)(N-3)}{2} = 2k(k-1) \ . 
\label{dif2a}
\eeq
\end{widetext}
This difference is also positive semidefinite, and positive-definite for $k \ge
2$, i.e., $N \ge 5$.  Hence, as these calculations show, a vacuum alignment
argument prefers the breaking pattern (\ref{symbreak_lam2neg}) for both even
and odd $N \ge 4$.  The special case $N=3$ has been analyzed above, and leads
to breaking of the SU(3) group to ${\rm SU}(2) \times {\rm U}(1)$, which is
also of the form (\ref{symbreak_lam2neg}) with $N=3$.

There are other symmetry-breaking patterns that could, {\it a priori}, occur.
SU($N$) could, in principle, break to a non-maximal subgroup, i.e., a subgroup
with rank smaller than the rank of SU($N$), namely $N-1$.  For example, in
principle SU(3) could, {\it a priori} break to U(1), SU(4) could break to ${\rm
SU}(2) \otimes {\rm U}(1)$, and so forth.  However, in the context of the Higgs
mechanism, these symmetry-breaking patterns do not occur as minima of the Higgs
potential, and in the dynamical symmetry breaking context, they are disfavored
by vacuum alignment arguments.

\subsection{SU($N$) Breaking with an Adjoint Higgs Field} 

First, we discuss the mechanism for breaking an SU($N$) gauge symmetry with a
Higgs field $\Phi$ in the adjoint representation \cite{lfli}. The components of
the Higgs field are denoted $\Phi^i_j$.  We impose a $\Phi \to -\Phi$
symmetry. Then the Higgs potential has the general form
\beq
V = \frac{\mu^2}{2} {\rm Tr}(\Phi^2) + 
\frac{\lambda_1}{4} [{\rm Tr}(\Phi^2)]^2 + 
\frac{\lambda_2}{4} {\rm Tr}(\Phi^4) \ , 
\label{vsu5}
\eeq
where we take $\mu^2 < 0$ to produce the symmetry breaking. Since $\Phi$ is a
hermitian matrix, it can be diagonalized by a unitary transformation.  It
follows that one can write
\beq
\Phi^i_j = \delta^i_j \phi_j  \quad ({\rm no \ sum \ on} \ j) \ , 
\label{phiform}
\eeq
for $1 \le i,j \le N$. Substituting Eq. (\ref{phiform}) into Eq. (\ref{vsu5})
gives
\beq
V = \frac{\mu^2}{2} \sum_{i=j}^N \phi_j^2 +
\frac{\lambda_1}{4}(\sum_{j=1}^N \phi_j^2)^2 +
\frac{\lambda_2}{4}\sum_{j=1}^N \phi_j^4 \ . 
\label{vphisimple}
\eeq
Since ${\rm Tr}(\Phi)=0$, the $\phi_j$ satisfy the condition
\beq
\sum_{j=1}^N \phi_j = 0 \ . 
\label{phisum}
\eeq
Hence, $\phi$ only involves $N-1$ independent fields, and $V$ only depends
on $N-1$ of the components $\phi_j$, which we take to be $\phi_j$ with 
$j=1,...,N-1$. Now
\beq
[{\rm Tr}(\Phi^2)]^2 \ge {\rm Tr}(\Phi^4) \ , 
\label{terminequality}
\eeq
as can be seen from the explicit expression for the difference, 
\beqs
& & [{\rm Tr}(\Phi^2)]^2 - {\rm Tr}(\Phi^4) =  \cr\cr
& & 2 \bigg [ \sum_{1 \le i < j \le N-1} \phi_i^2 \phi_j^2 + 
 \Big ( \sum_{i=1}^{N-1} \phi_i^2 \Big ) 
\Big ( \sum_{j=1}^{N-1} \phi_j \Big )^2 \bigg ] \cr\cr
& \ge & 0 \ . 
\label{term1minusterm2}
\eeqs
As noted above, if $N$ is equal to 2 or 3, then $[{\rm Tr}(\Phi^2)]^2 = 2{\rm
Tr}(\Phi^4)$, so that there is only one independent quartic term, and its
coefficient, $(1/4)[\lambda_1 + (\lambda_2/2)]$, must be positive. For $N \ge
4$, the two quartic terms are independent, and the condition that $V$ be
bounded from below requires that $\lambda_1 > 0$.

For $\lambda_2 >0$, it is again convenient to consider the cases of even $N=2k$
and odd $N=2k+1$ separately.  For $\lambda_2 > 0$ and even $N=2k$, $V$ is
minimized if the VEV of $\Phi$ has the form given by
\beq
\langle \phi_i \rangle =  \frac{v}{\sqrt{2N}} \times 
     \cases{ 1 & for $1 \le i \le k$ \cr
            -1 & for $k+1 \le i \le 2k$ } \ . 
\label{phivev_neven_lam2pos}
\eeq
The normalization of $v$ in Eq. (\ref{phivev_neven_lam2pos}) and in the
equations below is determined by the definition ${\rm Tr}(\Phi^2) = (1/2)v^2$,
analogous to the usual normalization condition ${\rm Tr}(T_aT_b) =
(1/2)\delta_{ab}$ for the generators of SU($N$).  At this minimum, one finds
\beq
v^2 = \frac{-2\mu^2}{\Big [\lambda_1+\frac{\lambda_2}{N} \Big ]} \ . 
\label{vvalue_neven_lam2pos}
\eeq
The VEV (\ref{phivev_neven_lam2pos}) breaks SU($2k$) according to 
(\ref{symbreak_neven_lam2pos}). The value of the potential at the minimum is
\beq
V_{min} = \frac{-\mu^4}{4\Big [\lambda_1+\frac{\lambda_2}{N} \Big ]} \ . 
\label{vpotmin_neven_lam2pos}
\eeq

For $\lambda_2 > 0$ and odd $N=2k+1$ with $k \ge 2$, $V$ is minimized 
if the VEV of $\Phi$ has the form 
\beqs
\langle \phi_i \rangle & = & v \left [\frac{k}{2(k+1)(2k+1)} \right ]^{1/2} 
\times 
\cr\cr & & 
\cr\cr
                  & \times & \cases{ 1 & for $1 \le i \le k+1$ \cr
   -\frac{k+1}{k} & for $k+2 \le i \le 2k+1$ } \ . 
\label{phivev_nodd_lam2pos}
\eeqs
(The special case $k=1$, i.e., $N=3$, was dealt with
above.) The minimization condition determines $v$ according to 
\beq
v^2 = \frac{-2\mu^2}{\bigg [ \lambda_1 + 
\Big ( \frac{N^2+3}{N(N+1)(N-1)} \Big )\lambda_2 \bigg ]} \ . 
\label{vvalue_nodd_lam2pos}
\eeq
This VEV (\ref{phivev_nodd_lam2pos}) breaks SU($2k+1$) according to 
(\ref{symbreak_nodd_lam2pos}). The value of the potential at the minimum is 
\beq
V_{min} = \frac{-\mu^4}{4 \bigg [ \lambda_1 + 
\Big ( \frac{N^2+3}{N(N+1)(N-1)} \Big )\lambda_2 \bigg ]} 
\label{vpotmin_nodd_lam2pos}
\eeq

It is possible for $\lambda_2$ to have a restricted range of negative values
\cite{negativerange}, 
\beq
-\bigg ( \frac{N(N-1)}{N^2-3N+3} \bigg ) \lambda_1 < \lambda_2 < 0 \ . 
\label{lam2range}
\eeq
For $\lambda_2$ in this range, $V$ is minimized if $\Phi$ has the VEV
\beq
\langle \phi_i \rangle = \frac{v}{\sqrt{2N(N-1)}} \times 
 \cases{ 1 & for $1 \le i \le N-1$ \cr
      -(N-1) & for $i=N$ } 
\label{phivev_lam2neg}
\eeq
where
\beq
v^2=\frac{-2\mu^2}{\bigg [\lambda_1 + 
\Big (\frac{N^2-3N+3}{N(N-1)} \Big )\lambda_2 \bigg ]} \ . 
\label{vsqvalue_lam2neg}
\eeq
The VEV (\ref{phivev_lam2neg}) breaks SU($N$) according to
Eq. (\ref{symbreak_lam2neg}). The value of $V$ at this minimum is 
\beq
V_{min} = \frac{-\mu^4}{4 \bigg [\lambda_1 + 
\Big (\frac{N^2-3N+3}{N(N-1)} \Big )\lambda_2 \bigg ]} \ . 
\label{vpotmin_lam2neg}
\eeq
Note that all three of the minimal values (\ref{vpotmin_neven_lam2pos}), 
(\ref{vpotmin_nodd_lam2pos}), and (\ref{vpotmin_lam2neg}) have the form
\beq
V_{min} = \frac{\mu^2 v^2}{8} 
\label{vminform}
\eeq
for the respective three values of $v^2$.  The lower limit on the allowed
negative range of $\lambda_2$ in Eq.  (\ref{lam2range}) is evident from
Eq. (\ref{vpotmin_lam2neg}), since in this equation $V_{min}  \to -\infty$ 
as $\lambda_2$ approaches this lower limit from above. The fact that 
$\lambda_2=0$ is the boundary between the two types of minima is evident from
the difference between the values of the minima for even $N$, 
\begin{widetext} 
\beq
V_{min, \ \lambda_2 > 0} - V_{min, \ \lambda_2 < 0} = 
= \frac{-(N-2)^2 \, \lambda_2 \, \mu^4 }
{4 \bigg [ \lambda_1 + \frac{\lambda_2}{N} \bigg ] 
\bigg [ N(N-1)\lambda_1 + (N^2-3N+3)\lambda_2 \bigg ]} 
\label{vmin_neven_lam2pos_minus_vmin_lam2neg}
\eeq
and for odd $N$, 
\beq
V_{min, \ \lambda_2 > 0} - V_{min, \ \lambda_2 < 0} = 
\frac{-N^3(N-1)(N-3) \, \lambda_2 \, \mu^4}
{4 \bigg [ N(N^2-1)\lambda_1 + (N^2+3)\lambda_2 \bigg ] 
\bigg [ N(N-1)\lambda_1 + (N^2-3N+3)\lambda_2 \bigg ] } \ . 
\label{vmin_nodd_lam2pos_minus_vmin_lam2neg}
\eeq
Both of these differences are proportional to $\lambda_2$, explicitly showing
the switch in global minimum as $\lambda_2$ reverses sign. 
\end{widetext}
The reason for the residual U(1) invariance in these symmetry-breaking patterns
obtained with a Higgs field $\Phi$ transforming according to the adjoint
representation of SU($N$) is that since $\Phi$ can be diagonalized, as noted
above, its VEV can be expressed as a linear combination of coefficients
multipled by the $N-1$ diagonal Cartan generators of SU($N$).  Indeed, without
loss of generality, one can define axes in ${\rm SU}(N)$ space so that it
points entirely along one such Cartan generator, which can be denoted as
$T_C$. Then $\exp(i \theta T_C)$ commutes with $\langle \Phi \rangle$, yielding
the U(1) invariance.

From the formulas for $\Delta o$, the number of broken generators for the
various symmetry-breaking patterns, one can immediately infer the number of
gauge bosons of SU($N$) that become massive.  Thus, for $\lambda_2 > 0$ and
even $N=2k$, the symmetry breaking (\ref{symbreak_neven_lam2pos}) involves the
breaking of $N^2/2=2k^2$ generators, so that of the $N^2-1$ (real) components
of $\Phi$, $N^2/2$ are absorbed to become the longitudinal components of the
gauge bosons corresponding to these broken generators, which pick up 
masses $\propto g v$.  The remaining $N^2/2-1=2k^2-1$ real components of $\Phi$
are physical Higgs bosons.  For $\lambda_2 > 0$ and odd $N=2k+1$, the symmetry
breaking (\ref{symbreak_nodd_lam2pos}) involves the breaking of
$(N^2-1)/2=2k(k+1)$ generators, so that of the $N^2-1$ (real) components of
$\Phi$, $(N^2-1)/2$ are absorbed to become the longitudinal components of the
gauge bosons corresponding to these broken generators. The remaining
$(N^2-1)/2=2k(k+1)$ real components of $\Phi$ are physical Higgs bosons.  For
$\lambda_2 < 0$, the symmetry breaking (\ref{symbreak_lam2neg}) involves the
breaking of $2(N-1)$ generators, and an equal number of Nambu-Goldstone bosons,
which are absorbed to become the longitudinal components of the gauge bosons in
the coset (\ref{coset_lam2neg}).  The remaining $(N-1)^2$ real components of
$\Phi$ are physical Higgs bosons.

\subsection{Dynamical Mechanism for ${\rm SU}(N)$ Breaking by an Adjoint 
Field} 

For the analysis of dynamical symmetry breaking of SU($N$) by an adjoint field,
we analyze a model of the form of Eq. (\ref{guv}), in which
\beq
G = {\rm SU}(N) \ , \quad G_b = {\rm SU}(N_b) \ . 
\label{sunmodel}
\eeq
For the fermions that transform under both SU($N$) and $G_b$ we use 
\beq
(\psi^i_j)^\alpha_L \ : \quad (N^2-1,N_b) \ , 
\label{psi}
\eeq
where here and below, $\alpha$ is the $G_b$ gauge index. Thus, we assign each
of the $N^2-1$ components of $(\psi^i_j)^\alpha_L$ to transform according to
the fundamental representation of SU($N_b$). The numbers in
parentheses in Eq. (\ref{psi}) are the dimensions of the representations with
respect to the factor groups in Eq. (\ref{sunmodel}). The $N_b$ copies
of fermions in the adjoint representation of SU($N$) contribute zero gauge
anomaly to SU($N$).  As stated earlier, these and the other fermions that we
include are taken to have zero Lagrangian masses since mass terms would violate
the full $G_{UV}$ symmetry, which is chiral.

The choice of the rest of the $G_b$-nonsinglet fermions in the model depends on
the value of $N_b$.  We first consider the possibility that $N_b=2$.  Now, $N$
is even $\Longleftrightarrow$ $N^2-1$ is odd.  The SU(2)$_b$ theory must have
an even number of chiral doublet fermions in order to avoid a global anomaly,
so if $N$ is odd, the $N^2-1$ $(\psi^i_j)^\alpha_L$ form an acceptable
SU(2)$_b$ fermion sector by themselves, while if $N$ is even, then we obtain an
acceptable fermion sector by adding an odd number of additional SU(2)$_b$
doublets.  We shall choose this odd number to be the minimal value, namely,
one, with the fermion
\beq
\omega^\alpha_L  \quad {\rm included \ for \ even } \ N \ . 
\label{omeganb2}
\eeq
For these two cases, the SU(2)$_b$ beta function has as its leading coefficient
\beq
b_1 = \cases{ \frac{1}{3}(23-N^2) \quad {\rm for} \ N \ {\rm odd} \cr\cr
              \frac{1}{3}(22-N^2) \quad {\rm for} \ N \ {\rm even} } 
\label{b1su2badjoint}
\eeq
The requirement that the SU(2)$_b$ theory be asymptotically free is thus that
$N < \sqrt{23}$ for odd $N$ and $N < \sqrt{22}$ for even $N$.  These amount to
the possibilities $N=3$ for odd $N$ and $N=2, \ 4$ for even $N$.  We have
already dealt with the case $N=3$ above, so here we focus on the case $N=4$.
As discussed in the introduction, these are necessary but not sufficient
conditions; we also must require that the fermion content of the SU(2)$_b$
theory is sufficiently small that as the reference energy scale $\mu$
decreases, the coupling $\alpha_b(\mu)$ will increase sufficiently so that the
SU(2)$_b$ gauge interaction will produce bilinear fermion condensates instead
of evolving in a chirally symmetric manner into the infrared.  For SU(2), the
critical number of Dirac fermions, $N_{f,cr}$, below which this condensation
will occur is estimated to be $N_{f,cr} \simeq 8$ \cite{chipt}.  Because
SU(2) has only (pseudo)real representations, we can rewrite the theory
with a given number of chiral Weyl doublets as a theory with half this number
of Dirac doublets.  For $N=4$, we would have $N^2=16$ chiral doublets, or eight
Dirac doublets, which is marginal.  Assuming that the SU(2)$_b$ sector does,
indeed, produce bilinear fermion condensates, these would occur in the most
attractive channel, which is $2 \times 2 \to 1$ in SU(2)$_b$.  These would have
either the form
\beq
\langle \epsilon_{\alpha \beta} [\psi_{a,L}^{\alpha \ T} C
  \psi_{b,L}^\beta]_{as} \rangle
\label{su4b_psipsi_condensate}
\eeq
or the form 
\beq
\langle \epsilon_{\alpha \beta} \psi_{a,L}^{\alpha \ T} C \omega^\beta_L
\rangle \ , 
\label{su4b_psiomega_condensate}
\eeq
where in Eq. (\ref{su4b_psipsi_condensate}) the symbol $[...]_{as}$ means an
antisymmetric SU(4) combination of the two adjoint fermion fields. In both
cases, the condensate thus transforms as an adjoint of SU(4).  A vacuum
alignment argument implies that the condensates form in such a way as to
preserve the largest subgroup in SU(4).  The order of the 
subgroup ${\rm SU}(3) \otimes {\rm U}(1)$ is 9, which is greater than the 
order of the subgroup ${\rm SU}(2) \otimes {\rm SU}(2) \otimes {\rm U}(1)$, 
which is 7.  Hence, from a vacuum alignment argument, one may infer that the
condensate is proportional to the SU(4)
generator $T_{15}=(2\sqrt{6})^{-1} \, {\rm diag}(1,1,1,-3)$, 
leading to the $N=4$ special case of the symmetry-pattern pattern 
(\ref{symbreak_lam2neg}). 

We next consider possible values $N_b \ge 3$ for the gauge group symmetry
SU($N_b$) responsible for the dynamical breaking of SU($N$). In this case, for
the rest of the $G_b$-nonsinglet fermions we choose
\beq
\omega^\alpha_{p,L} \ : \quad (N^2-1)(1,\bar N_b) \ , 
\label{omega}
\eeq
where the notation $\bar N_b$ means the conjugate fundamental representation
and here the copy number takes on the values $1 \le p \le N^2-1$.  This ensures
that the SU($N_b$) theory has zero gauge anomaly. With the fermions (\ref{psi})
and (\ref{omega}) (and with the SU($N$) interaction taken as negligibly weak),
the SU($N_b$) theory is vectorlike. This is in accord with one of the
conditions that we imposed above, which guarantees that the $G_b$ symmetry does
not self-break when it becomes strongly coupled.  Expressed in manifestly
vectorial form, it has $N^2-1$ Dirac fermions transforming according to the
fundamental representation of SU($N_b$).

The beta function for the SU($N_b$) coupling has leading coefficient
\beq
(b_1)_{SU(N_b)} = \frac{1}{3}[11N_b-2(N^2-1)] \ . 
\label{b1gb}
\eeq
The requirement that the SU($N_b)$ theory be asymptotically free is thus
\beq
N_b > \frac{2(N^2-1)}{11} \ . 
\label{nbaf}
\eeq
As noted above, this is a necessary, but not sufficient, condition for the
SU($N_b$) theory to produce the requisite condensates.  We must also require
that, for a given value of $N_b$, the fermion content of the SU($N_b$) sector
must be small enough so that as the theory evolves down in energy scale, it
produces condensates instead of evolving into the infrared in a chirally
symmetric (conformal) manner.  For a vectorial asymptotically free SU($N$)
gauge theory with $N_f$ copies of Dirac fermions (with zero Lagrangian masses)
in the fundamental representation, if $N_f$ is smaller than a critical value,
$N_{f,cr}$, then as the reference scale decreases from large values, the
coupling will eventually grow large enough to form condensates which
generically break the global chiral symmetry.  In contrast, if $N_f > N_{f,cr}$
then the theory will evolve from the ultraviolet to the infared without any
spontaneous chiral symmetry breaking, yielding conformal behavior.  A combined
analysis of the beta function and solutions of the Dyson-Schwinger equation for
the fermion propagator in the approximation of one-gluon exchange yields the
result \cite{chipt}
\beq
N_{f,cr} = \frac{2N_b(50N_b^2-33)}{5(5N_b^2-3)} \ .
\label{nfcr}
\eeq
Although the Dyson-Schwinger analysis does not directly incorporate either
effects of confinement or instantons, it has been shown that these two effects
affect $N_{f,cr}$ in opposite ways, so that neglecting both of them can still
yield a reasonably accurate result \cite{lmax}.  Recent lattice simulations of
SU(3) gauge theory with variable numbers $N_f$ of light fermions in the
fundamental representation are (taking account of theoretical uncertainties in
both Eq. (\ref{nfcr}) and the lattice work) broadly consistent with
Eq. (\ref{nfcr}) \cite{lgt}.  Although this does not test the prediction for
$N_b \ne 3$, it makes it plausible that this prediction could also be
reasonably accurate. For these values of $N_b$, Eq. (\ref{nfcr}) rapidly
approaches the asymptotic large-$N_b$ form $N_{f,cr} \simeq 4N_b$.  We thus
require that $N_b$ is sufficiently large that the SU($N_b$) theory with its
$N_f=N^2-1$ Dirac fermions will exhibit spontaneous chiral symmetry breaking
and confinement instead of evolving down in energy in a chirally symmetric
non-Abelian Coulomb (conformal) phase.  Using the prediction of
Eq. (\ref{nfcr}), we thus obtain the lower bound $N_{f,cr} \simeq 4N_b >
N^2-1$, i.e.,
\beq
N_b > \frac{(N^2-1)}{4} \ . 
\label{nbllower}
\eeq

With the fermion content as specified via Eqs. (\ref{psi}) and (\ref{omega}),
and in the approximation that one turns off the SU($N$) gauge interaction, the
SU($N_b$) sector has a classical global symmetry of the form ${\rm
U}(N^2-1)_{\psi} \otimes {\rm U}(N^2-1)_{\omega}$, or equivalently, ${\rm
SU}(N^2-1)_{\psi} \otimes {\rm SU}(N^2-1)_{\omega} \otimes {\rm U}(1)_{\psi}
\otimes {\rm U}(1)_{\omega}$, where the subscripts indicate which fields are
involved in the respective symmetry transformations.  Both the U(1)$_\psi$ and
U(1)$_\omega$ are broken by SU($N_b$) instantons, but the linear combination
corresponding to the difference of the currents for the $\psi$ and $\omega$
fields is conserved in the presence of instantons.  We will denote this
symmetry as U(1)$'$.  The actual (non-anomalous) global symmetry of the $G_b$
theory at the high scale is thus
\beq
{\rm SU}(N^2-1)_{\psi} \otimes {\rm SU}(N^2-1)_{\omega} \otimes
{\rm U}(1)' \ . 
\label{sun_globalsym}
\eeq
We comment on this further below. 

Now we turn on the SU($N$) gauge interaction.  This explicitly breaks the
above global chiral symmetry. However, just as the breaking of chiral ${\rm
SU}(2)_L \otimes {\rm SU}(2)_R$ symmetry in QCD by electroweak interactions is
weak, so also here this breaking is weak, since $\alpha_G$ is small at the
scale $\Lambda_b$. We can fix the initial value of $\alpha_b(\mu)$ at a high
value of $\mu$ so that as $\mu$ decreases to the scale $\Lambda_b$, this
coupling grows sufficiently large to produce bilinear fermion condensates.
These condensates will occur in the most attractive channel, which, for the
above fermion content, is $N_b \times \bar N_b \to 1$. In general, these 
condensates would be of the form $\langle \psi_{a,L}^{\alpha \ T} C
\omega_{p,\alpha,L} \rangle$.  A vacuum alignment argument implies that these
condensates will form in a manner such as to preserve the largest residual
gauge symmetry. We regard this implication as very plausible, but add the
obvious caveat that one must remember the theoretical uncertainties that are
present in such a strongly coupled theory. Combining this implication from the
vacuum alignment argument with our discussion above, we infer that the
symmetry-breaking pattern is that SU($N$) breaks to the maximal subgroup 
${\rm SU}(N-1) \otimes {\rm U}(1)$ as in Eq. (\ref{symbreak_lam2neg}), so that
the condensate would have the form
\beq
\langle \psi_{a,L}^{\alpha \ T} C \omega_{p,\alpha,L} \rangle  \quad {\rm with}
\ a = N^2-1 \ . 
\label{symcond}
\eeq
That is, it would transform like the last of the generators in the Cartan
algebra of SU($N$), 
\beqs
& & (T_{a=N^2-1})^i_j = \cr\cr
& & \frac{1}{\sqrt{2N(N-1)}}\,\delta^i_j \cases{ 1 & for $1 \le i \le N-1$
                                             \cr-(N-1)  & for $i = N$ } \cr\cr
& & 
\label{tlast}
\eeqs

Here we emphasize an important contrast between this dynamical symmetry
breaking mechanism and the Higgs mechanism.  In our introductory discussion
above, we have already noted a number of the differences between the Higgs
mechanism and a dynamical mechanism for breaking a gauge symmetry.  Among other
differences, for example, the Higgs mechanism leads to the appearance of at
least one physical pointlike Higgs field, whereas a dynamical mechanism does
not yield such a particle (although it may yield composite $J=0$ bound states).
Furthermore, if one uses a Higgs mechanism to break SU($N$), then by
appropriate choices of the parameters, one can guarantee that the minimim of
the potential occurs for a Higgs VEV of the form (\ref{phivev_neven_lam2pos})
or (\ref{phivev_nodd_lam2pos}), so that the symmetry breaking is of the type
(\ref{symbreak_neven_lam2pos}) or (\ref{symbreak_nodd_lam2pos}), rather than
(\ref{symbreak_lam2neg}).  However, in the dynamical approach to SU($N$)
breaking, once one specifies the gauge and fermion content, there are no free
parameters, and the theory is, in principle, completely predictive. Although
the dynamical symmetry-breaking mechanism involves a strongly coupled gauge
sector, one can use most attractive channel criteria and vacuum alignment
arguments to make a plausible inference about what form the bilinear fermion
condensate will take, namely, as discussed above, the form that preserves the
largest residual symmetry, ${\rm SU}(N-1) \otimes {\rm U}(1)$.  These MAC and
vacuum alignment properties would be manifest if one were to explicitly
calculate the effective potential for the composite operator represented by the
condensate, along the lines of Ref. \cite{cjt}.  In this context, one may
recall that the Higgs potential was partially motivated by the original
Ginzburg-Landau free energy functional in phenomenological models of
superconductivity, and retrospectively, from the perspective of the
Bardeen-Cooper-Schrieffer theory and the Cooper pair condensate, one may view
the Ginzburg-Landau free energy functional as an approximate way to represent
the physics of this Cooper pair condensate.  This is, of course, not a precise
isomorphism, but only a partial correspondence.  As recalled above, there are
important differences between a Higgs and dynamical mechanism for breaking a
gauge symmetry.  To the extent that one may regard a Higgs potential as
embodying some of the same physics as an effective potential for a composite
operator represented by bilinear fermion condensate(s), one may observe that
the pattern of symmetry breaking inferred from the dynamical approach makes
definite predictions for the coefficients in the corresponding Higgs
potential. First, because the Lagrangian in the dynamical model is invariant
under the separate global transformations $\psi_{a,L}^\alpha \to -
\psi_{a,L}^\alpha$ and $\omega^\alpha_{p,L} \to - \omega^\alpha_{p,L}$, it
follows that an analogous effective potential for the condensate
(\ref{symcond}) should not contain odd powers of this condensate.  Our
dynamical model for the symmetry breaking of an SU($N$) gauge theory using
fermions transforming as an adjoint representation of SU($N$) then predicts
that in a corresponding Higgs approach, in order to obtain the same pattern of
symmetry breaking, the coefficients of the Higgs potential should have the
following properties: (i) $\mu^2 < 0$, for symmetry breaking; (ii) $\lambda_2 <
0$, yielding the specific symmetry-breaking pattern (\ref{symbreak_lam2neg});
and the stability properties that (iii) $\lambda_1 > 0$ and (iv) $\lambda_2$
satisfies the lower bound in Eq. (\ref{lam2range}).

With the symmetry-breaking pattern as given by (\ref{symbreak_lam2neg}), there
are then $2(N-1)$ broken generators of SU($N$), and Nambu-Goldstone modes
formed from the fermion condensates are absorbed by the gauge bosons
corresponding to these broken generators, forming the longitudinal components
of the resultant massive vector bosons.  These masses are of order
$g\Lambda_b$. This is reminiscent of the process whereby Nambu-Goldstone modes
in the technicolor mechanism for electroweak symmetry breaking are absorbed to
give the $W^\pm$ and $Z$ bosons their masses.  The SU($N_b$)-nonsinglet
fermions involved in the condensate (\ref{symcond}) gain dynamical masses of
order $\Lambda_b$ and are integrated out of the low-energy effective theory
that is operative at scales $\mu$ below $\Lambda_b$.  Since, by construction,
the SU($N_b$) theory confines, the spectrum of the SU($N_b$) theory includes a
set of SU($N_b$)-singlet mesons, baryons, and glueballs that form at the scale
$\Lambda_b$.

\section{Remarks on Other Directions of Study }

We comment here on some other related directions of study that could be
interesting to pursue. One could construct models with dynamical symmetry
breaking of other gauge symmetries and compare results with those obtained via
Higgs scenarios.  An example of this would be models with extended electroweak
gauge groups such as $G = {\rm SU}(3)_c \otimes {\rm SU}(2)_L \otimes {\rm
SU}(2)_R \otimes {\rm U}(1)_{B-L}$ and $G = {\rm SU}(4) \otimes {\rm SU}(2)_L
\otimes {\rm SU}(2)_R$, for which dynamical mechanisms were presented in
Ref. \cite{lrs}.  In a more abstract direction, one could consider groups such
as $G={\rm SO}(N)$.  One could also study the breaking of SU($N$) by fields
transforming according to representations other than the fundamental and
adjoint, such as the rank-2 symmetric and antisymmetric tensor
representations. 

One could also study situations in which the $G$ gauge interaction is not
weakly coupled at the scale $\Lambda_b$ where the $G_b$ interaction becomes
strongly coupled, so that there is generically a combination of self-breaking
of $G$ and induced breaking of $G$ by $G_b$.  Indeed, in reasonably
ultra-violet-complete extended technicolor (ETC) theories, the sequential
breaking of the ETC gauge symmetry down to the residual exact technicolor
symmetry typically involves both self-breaking of ETC, which is a strongly
coupled, chiral gauge symmetry, and induced breaking by an auxiliary gauge
interaction called hypercolor in \cite{uvetc}.  A similar statement applies to
ultraviolet completions of topcolor-assisted technicolor models that include
the necessary additional gauge interactions to produce the required symmetry
breakings \cite{tcrev,tc2uv}.

Although our study is primarily intended as a comparison of gauge symmetry
breaking by dynamical and Higgs mechanisms in a general field theoretic
context, it is appropriate to address the question of possible dynamical
symmetry breaking of a grand unified symmetry.  We recall that there has long
been interest in grand unified theories (GUTs) which embed the three factor
groups of $G_{SM}$ in a single group, since this would unify quarks and
leptons, predict the ratios of the three SM gauge couplings, and quantize
electric charge \cite{gg}-\cite{rabypdg}.  Much work on GUTs has been done in a
supersymmetric context, since supersymmetry remedies the gauge hierarchy
problem of the Standard Model and since the MSSM naturally yields gauge
coupling unification.  There have also been studies of the question of whether
some type of grand unification could feasibly be achieved in a theoretical
context involving dynamical electroweak symmetry breaking \cite{etcgut}.  It is
natural to ask whether one could use induced dynamical breaking of a GUT gauge
symmetry such as SU(5) or SO(10), which is weakly coupled at the GUT scale,
$M_{GUT}$, using a (vectorial non-Abelian, asymptotically free) $G_b$ gauge
interaction that becomes strongly coupled at this scale.  One could, of course,
argue that such an approach differs from the original purpose of the grand
unification, which was to obtain an ultraviolet-scale theory with only a single
gauge group and gauge coupling.  Indeed, such an induced GUT symmetry breaking
scenario appears problematic, since in order to produce the requisite bilinear
fermion condensates, the $G_b$ interaction would necessarily have to confine,
and this would generically lead to stable $G_b$-singlet baryons with masses of
order $M_{GUT}$.  With plausible estimates for the relevant reaction cross
sections, one finds that these $G_b$-singlet baryons would contribute far too
much to the dark matter in the universe \cite{rsu}.  Interestingly, even if a
dynamical approach to breaking a GUT symmetry were not excluded by its
production of excessive dark matter, it would predict that a GUT group such as
SU(5) would preferentially break to ${\rm SU}(4) \otimes {\rm U}(1)$ rather
than the SM group, ${\rm SU}(3)_c \otimes {\rm SU}(2)_L \otimes {\rm U}(1)_Y$.
In the conventional SU(5) GUT, the latter breaking to $G_{SM}$ is obtained by a
Higgs mechanism with a Higgs field transforming as the adjoint representation
\cite{gg}. Modern GUT theories also make use of string-inspired mechanisms for
the GUT gauge symmetry breaking, including higher-dimension operators and
Wilson lines \cite{rabypdg}.

\section{Conclusions}

In conclusion, in this paper we have constructed and analyzed theories with a
gauge symmetry in the ultraviolet of the form $G \otimes G_b$, in which the
vectorial, asymptotically free $G_b$ gauge interaction becomes strongly coupled
at a scale where the $G$ interaction is weakly coupled and produces bilinear
fermion condensates that dynamically break the $G$ symmetry.  We have compared
the results to those obtained with a Higgs mechanism.  There are many
interesting contrasting properties of these two approaches to breaking a gauge
symmetry.  The Higgs mechanism is perturbative, and one has the freedom, by
appropriate choices of parameters in the Higgs potential, to determine whether
and, in general, how the symmetry breaks.  In contrast, the dynamical approach
is arguably more predictive, in the sense that, provided that one has chosen
the gauge and field content of the $G_b$ sector appropriately, there are no
free parameters to vary; the $G_b$ gauge interaction will confine and produce
fermion condensates that break the $G$ symmetry.  Most attractive channel and
vacuum alignment arguments provide a plausible guide to enable one to infer
which channel(s) have fermion condensation, and what the form of this
condensation is, thereby predicting the resultant pattern of symmetry breaking.
In the dynamical models that we have constructed, we produce this breaking by
introducing fermions that are nonsinglets under both $G$ and $G_b$.  In the
course of our analysis, we have discussed how the gauge symmetry $G$ can be
broken not just for the case where it is chiral (as in electroweak symmetry
breaking), but also for the case where it is vectorial.  We have compared Higgs
and dynamical mechanisms for breaking SU(3) via a Higgs field or condensate
transforming according to the fundamental or adjoint representation.  We have
also carried such an analogous study for SU($N$) with $N \ge 4$.  Our present
study helps to elucidate the differences between the Higgs and dynamical
mechanisms for breaking a gauge symmetry. We believe such theoretical studies
are useful since it is still an open question what mechanism is responsible for
breaking electroweak gauge symmetry or a grand unified symmetry.

\bigskip
\bigskip

Acknowledgments: This research was partially supported by the grant
NSF-PHY-06-53342.

\section{Appendix I} 

Here we define some notation used in the text. For a gauge group $G_j$ we
denote the running gauge coupling as $g_j(\mu)$, where $\mu$ is the Euclidean
reference momentum, and we denote $\alpha_j(\mu)=g_j(\mu)^2/(4\pi)$. The beta
function is $\beta_{G_j} = dg_j/dt$, where $dt = d\ln \mu$.  We write
\beq
\frac{d\alpha_j}{dt} = - \frac{\alpha_j^2}{2\pi} \left [ b_1 + 
\frac{b_2 \, \alpha_j}{4\pi} + 
O(\alpha_j^3) \right ] 
\label{beta}
\eeq
where the first two coefficients, $b_1$ and $b_2$, are scheme-independent.  
For a representation $R$ of a Lie group $G$, the quadratic Casimir
invariant $C_2(R)$ is defined by 
$\sum_{a=1}^{order(G)} \sum_{j=1}^{dim(R)} (T_a)_{ij}(T_a)_{jk} = 
C_2(R)\delta_{ik}$.


\begin{thebibliography}{99}
\bibitem{triviality}
%
Indeed, if one were to envision fixing $\lambda$ at an arbitrarily high scale
$\mu$, then the value of $\lambda$ at a low scale would approach zero, 
the property known as triviality.
\bibitem{deltac2}
%
In general, if one has fermions transforming according to the representations
$R_1$ and $R_2$ of a gauge group, then a measure of the attractiveness of the
condensation channel $R_1 \times R_2 \to R_{cond}$ is $\Delta C_2 =
C_2(R_1)+C_2(R_2)-C_2(R_{cond})$, where $C_2(R)$ is the quadratic Casimir
invariant for the representation $R$.
\bibitem{selfbreak}
%
This is to be contrasted with the situation in a strongly coupled chiral gauge
theory, where the associated gauge symmetry does self-break \cite{rds}.
\bibitem{negativerange}
%
Although some of our discussion in this section is review, the bounded range
for negative $\lambda_2$ in Eq. (\ref{lam2range}) and its consequences do not
seem to have been noted before in the literature.
\end{thebibliography}
\end{document}